\documentclass[twocolumn,superscriptaddress,reprint,floatfix,amsmath,amssymb]{revtex4-1} 

\usepackage{graphicx}
\usepackage{dcolumn}
\usepackage{bm}
\usepackage{amssymb}
\usepackage{epsfig}
\usepackage{amsmath}
\usepackage{amsfonts}
\usepackage{mathptmx}
\usepackage{dcolumn}
\usepackage{eucal}
\usepackage{color}
\usepackage[colorlinks,linkcolor=blue,citecolor=blue]{hyperref}

\begin{document}
\title{\textbf{The $\omega$-SQUIPT: phase-engineering of Josephson topological materials}}
\author{E. Strambini}
\email{Equal contributor}
\affiliation{NEST Istituto
Nanoscienze-CNR  and Scuola Normale Superiore, I-56127 Pisa, Italy}
\author{S. D'Ambrosio}
\email{Equal contributor}
\affiliation{NEST Istituto
Nanoscienze-CNR  and Scuola Normale Superiore, I-56127 Pisa, Italy}
\author{F. Vischi}
\affiliation{NEST Istituto
Nanoscienze-CNR  and Scuola Normale Superiore, I-56127 Pisa, Italy}
\author{F. S. Bergeret}
\affiliation{Centro de Fisica de Materiales (CFM-MPC), Centro Mixto CSIC-UPV/EHU,
Manuel de Lardizabal 5, E-20018 San Sebastian, Spain}
\affiliation{Donostia International Physics Center (DIPC),
Manuel de Lardizabal 5, E-20018 San Sebastian, Spain}
\author{Yu. V. Nazarov}
\affiliation{Kavli Institute of Nanoscience, Delft University of Technology,Lorentzweg 1,2628 CJ, Delft, The Netherlands}
\author {F. Giazotto}
\email{francesco.giazotto@sns.it}
\affiliation{NEST Istituto
Nanoscienze-CNR  and Scuola Normale Superiore, I-56127 Pisa, Italy}
\maketitle
\textbf{Multi-terminal superconducting Josephson junctions based on the proximity effect offer the bright opportunity to tailor non trivial quantum states in nanoscale weak-links. 
These structures can realize exotic 
topologies in multidimensions~\cite{Riwar_Multi-terminal_2015} as, for example, artificial topological superconductors able to support Majorana bound states \cite{Mourik_signatures_2012,Sau_generic_2010}, and pave the way to emerging quantum technologies~\cite{Padurariu_Closing_2015,yokoyama_singularities_2015,van_Heck_single_2014,Rech_Proposal_2014} and future quantum information schemes \cite{padurariu_Spin_2012}.
Here, we report the first realization of a three-terminal  Josephson interferometer based on  a proximized nanosized weak-link.
Our tunneling spectroscopy measurements reveal transitions between gapped (i.e., insulating) and gapless (i.e., conducting) states, those being controlled by the phase configuration of the three superconducting leads connected to the junction.
We demonstrate the \emph{topological} nature of these transitions: a gapless state necessarily occurs between two gapped states of different topological index, very much like the interface between two insulators of different topology is necessarily conducting~\cite{Qi_Topological_2011}. 
The topological numbers characterizing such gapped states are given by superconducting phase windings over the two loops forming the Josephson interferometer. 
Since these gapped states cannot be transformed to one another continuously withouth passing through a gapless condition, these are topologically \emph{protected}.
Our observation of the gapless state is pivotal for enabling phase engineering of more sophisticated artificial topological materials realizing Weyl points or the anomalous Josephson effect \cite{Padurariu_Closing_2015,yokoyama_singularities_2015,Riwar_Multi-terminal_2015,van_Heck_single_2014,Rech_Proposal_2014,Pfeffer_Subgap_2014}.}

When two superconductors (S) are coupled through a normal metal (N), they realize a Josephson junction (JJ) and superconducting correlations are induced in the N region due to \emph{proximity effect}~\cite{de_gennes_superconductivity_1966,Giazotto_Superconducting_2010,Meschke_Tunnel_2011,le_sueur_phase_2008,Petrashov_Phase_1995,Gueron_Superconducting_1996}.
As a consequence, the N metal acquires genuine superconducting-like properties such as the ability to sustain a supercurrent, and a gap in the density of states (DoS) whose amplitude can be controlled by the macroscopic phase difference between the S leads~\cite{Giazotto_Superconducting_2010,Giazotto_Hybrid_2011,Ronzani_Highly_2014,dambrosio_Normal_2015}.
This process therefore enables the N region to possess a character ranging from insulating-like (gapped state) to conducting-like (gapless state) \cite{le_sueur_phase_2008,Giazotto_Hybrid_2011}.
\begin{figure}[t!]
\includegraphics[width=\columnwidth]{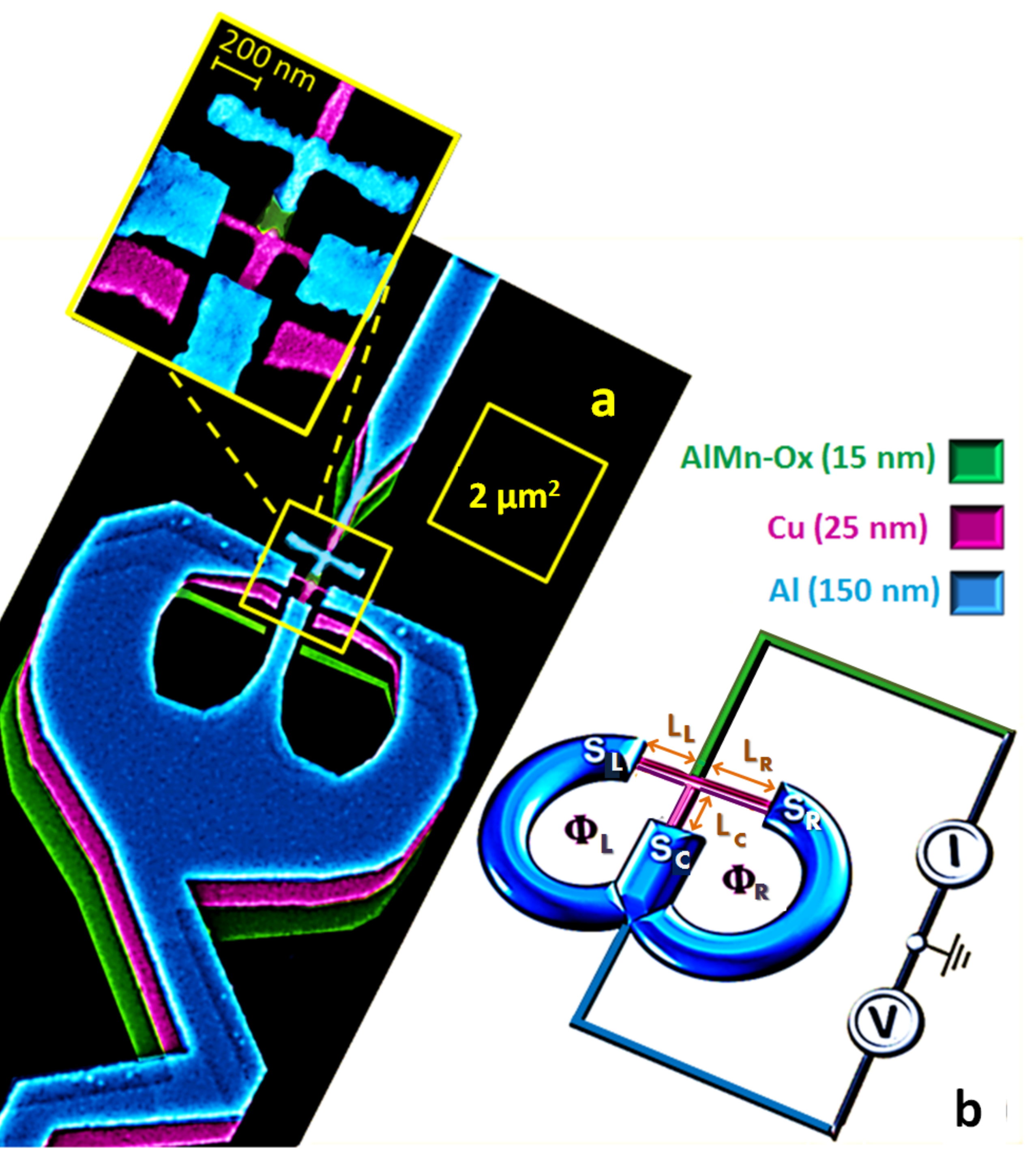}
\caption{
\label{sketch}
\textbf{The $\omega$-SQUIPT: a three-terminal double-loop Josephson interferometer based on proximity effect.}
\textbf{a},~Pseudo-color tilted scanning electron micrograph of a typical $\omega$-SQUIPT. 
The pseudo-color blow-up on the top of the figure highlights the core of the interferometer: a nanosized  T-shaped proximized Cu weak-link (magenta) in clean metallic contact with two Al superconducting loops (blue). The area of each loop is around $\sim$~2~$\mu$m$^{2}$. 
The central part of the weak-link is tunnel coupled to a $\sim 100$-nm-wide Al$_{0.98}$Mn$_{0.02}$ normal metal probe (green).
The structure
replicas resulting from the shadow-mask evaporation process are visible.
\textbf{b},~Scheme of the measurement setup. The current flowing through the circuit is indicated by $I$, and $V$ is the voltage drop across the interferometer. 
$\Phi_{L}$ and $\Phi_{R}$ represent the two magnetic fluxes piercing the left and right loop, respectively, whereas $L_{L}$, $L_{C}$, and $L_{R}$ denote the lengths of the three arms of the weak-link.
}
\end{figure}
\begin{figure*}[t!]
\includegraphics[width=0.98\textwidth]{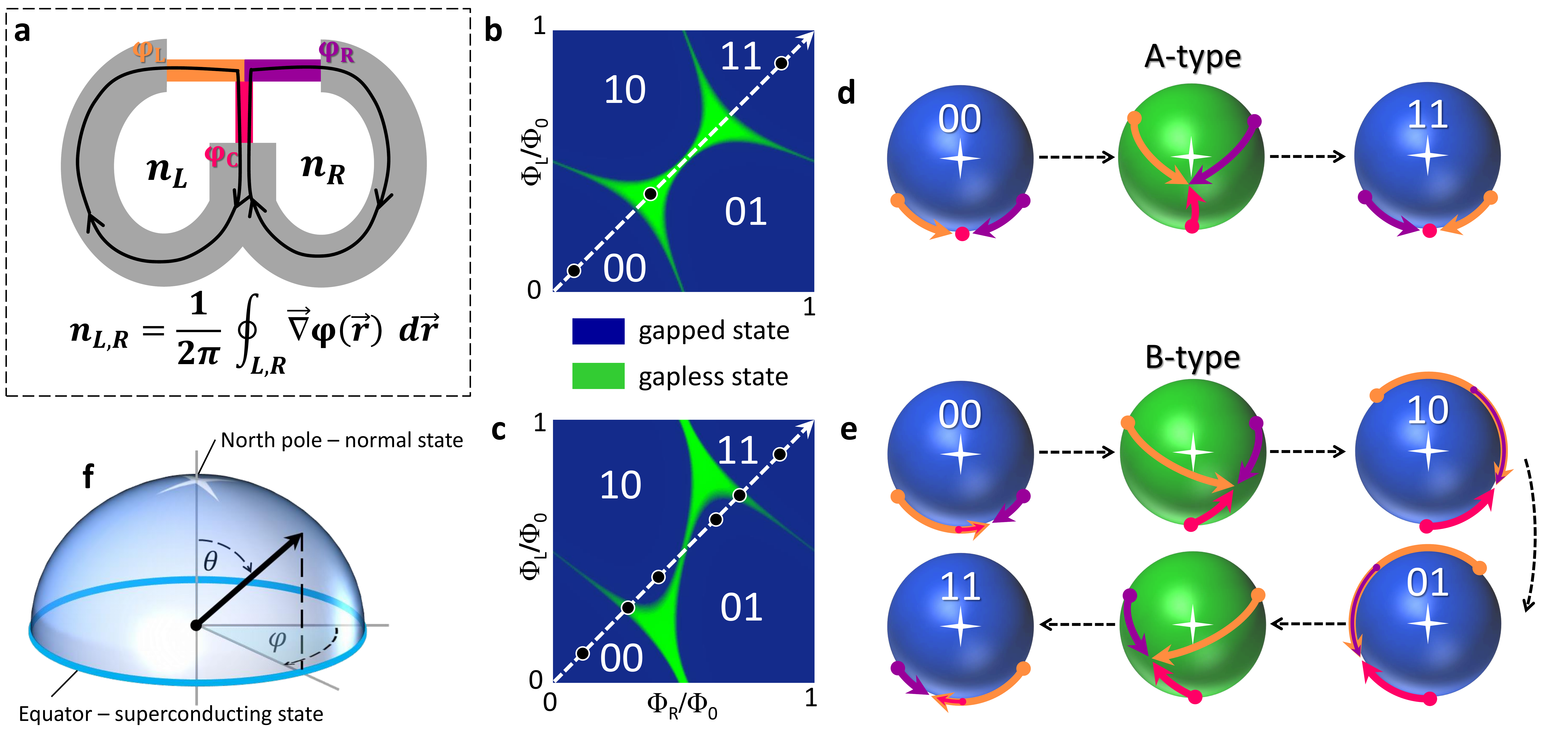}
\caption{
\label{Topology}
\textbf{Topological classes of the $\omega$-SQUIPT.}
\textbf{a},~Sketch of the double-loop interferometer with the geometric paths used to determine the topological index ($n_L,n_R$). These are defined by the closed line integral of the gradient of the S phase $\varphi(r)$ along the left ($n_L$) and right ($n_R$) loop. $\varphi_L$, $\varphi_C$, and $\varphi_R$ are the three superconducting phases at the interfaces with the T-shaped weak-link. 
\textbf{b},\textbf{c},~DoS at the Fermi energy calculated for an interferometer representing the A-type $\omega$-SQUIPT (\textbf{b}),  and for the B-type one (\textbf{c}). The blue areas indicate the insulating (gapped) states classified by the topological index in \textbf{a}. 
\textbf{d},\textbf{e}~Vectorial representation of the evolution of the topological state along the line defined by $\Phi_L=\Phi_R$ accessible in our experiment. 
The states at the three S leads lie at the equator of the unitary hemisphere (\textbf{f}), as well as the radial vector pointing at the intersection among the three arrows representing the weak-link in a gapped state.
Between each distinct topological configuration this quantum state necessarily evolves through a gapless state deviating from the equator towards the North pole represented by a cross on the  top of the hemisphere.
The \emph{latitudinal} angle ($\theta$) describes the degree of "superconductivity" of the weak-link whereas the \emph{longitudinal} angle ($\varphi$) represents the superconducting phase. 
}
\end{figure*}
Although two-terminal JJs based on proximity effect have been at the focus of an intense research for several years~\cite{Giazotto_Superconducting_2010,Giazotto_Hybrid_2011,Ronzani_Highly_2014,dambrosio_Normal_2015,de_gennes_superconductivity_1966,Meschke_Tunnel_2011,le_sueur_phase_2008,Josephson_Possible_1962,Petrashov_Phase_1995,Belzig_Negative_2002,Gueron_Superconducting_1996}, 
three-terminal junctions have been only recently realized to investigate the physics of multiple crossed Andreev reflections existing in metallic weak-links~\cite{Pfeffer_Subgap_2014}.
Yet, the impact and  control over three superconducting phases acting on a nanosized N region have never been explored so far despite recent predictions  of using multi-terminal JJs for tailoring and controlling exotic quantum states.
\cite{Padurariu_Closing_2015,van_Heck_single_2014,Riwar_Multi-terminal_2015,Pfeffer_Subgap_2014}.
Indeed, multi-terminal ($>$2) JJs allow the spin-orbit interaction to affect substantially the Andreev bound levels enabling the manipulation of electrons in single fermionic states~\cite{van_Heck_single_2014}. This is required for spin qubits~\cite{Loss_Quantum_1998} and spintronic applications~\cite{wolf_spintronics:_2001}, and can provide an alternative route towards the spooky Majorana bound states of topological superconductors \cite{Beenakker_Search_2013}, or the fundamental physics of Weyl singularities accessible in four or more terminal JJs~\cite{yokoyama_singularities_2015}.

\begin{figure*}[t!]
\includegraphics[width=0.98\textwidth]{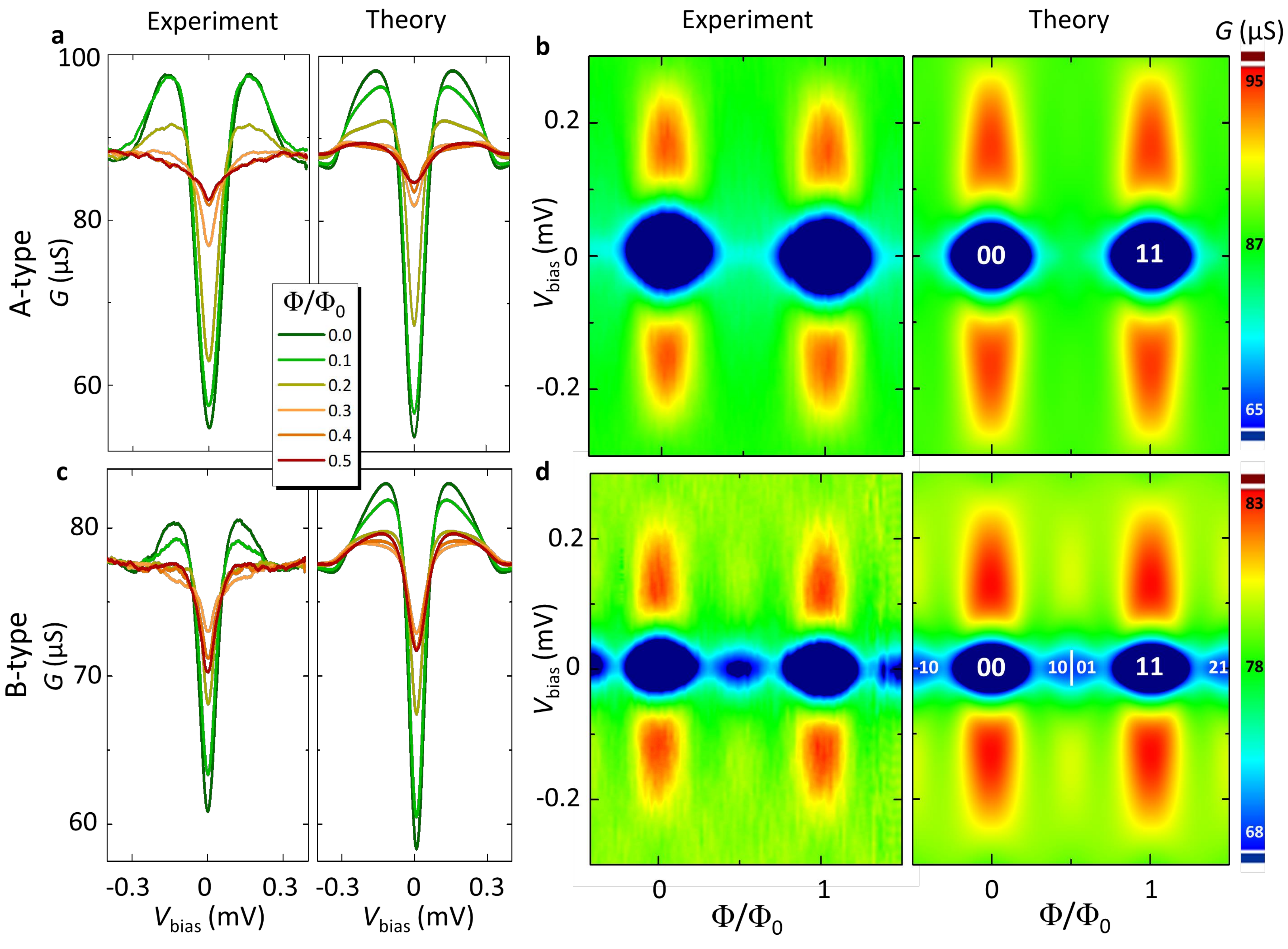}
\caption{
\label{TheEffect}
\textbf{Low-temperature magnetic-flux behavior of the two types of $\omega$-SQUIPTs.
}
\textbf{a},\textbf{c},~Comparison between the tunneling conductance $G$ vs bias voltage $V_{\text{bias}}$ measured at 30~mK (left panels) and calculated (right panels) for  selected values of the magnetic flux $\Phi$ ($\Phi_L=\Phi_R=\Phi$) for the A-type (\textbf{a}) and B-type (\textbf{c}) $\omega$-SQUIPT. 
Both the experimental and theoretical conductance spectra include a series resistance of $\sim 4.75$k$\Omega$ which stems from the measurement setup. The normal-state resistance $R_T$ of the A(B)-type interferometer is $\sim 6.5$k$\Omega$ ($\sim 8$k$\Omega$).
\textbf{b},\textbf{d},~ Color plot of measured (left panels) and calculated (right panels) $G$ vs $V_{\text{bias}}$ and $\Phi$ characteristics showing a \emph{gapless} regime (green region) for $\Phi_0/4 \lesssim \Phi \lesssim 3/4 \Phi_0$  in the A-type interferometer (\textbf{b}), and the presence of a second small minigap within the same flux interval for the B-type  $\omega$-SQUIPT (\textbf{d}). For each gapped state (blue areas) the corresponding topological index is shown. $\Phi_0=2.067\times 10^{-15}$ Wb is the flux quantum.
}
\end{figure*}
\begin{figure*}[t!]
\includegraphics[width=0.92\textwidth]{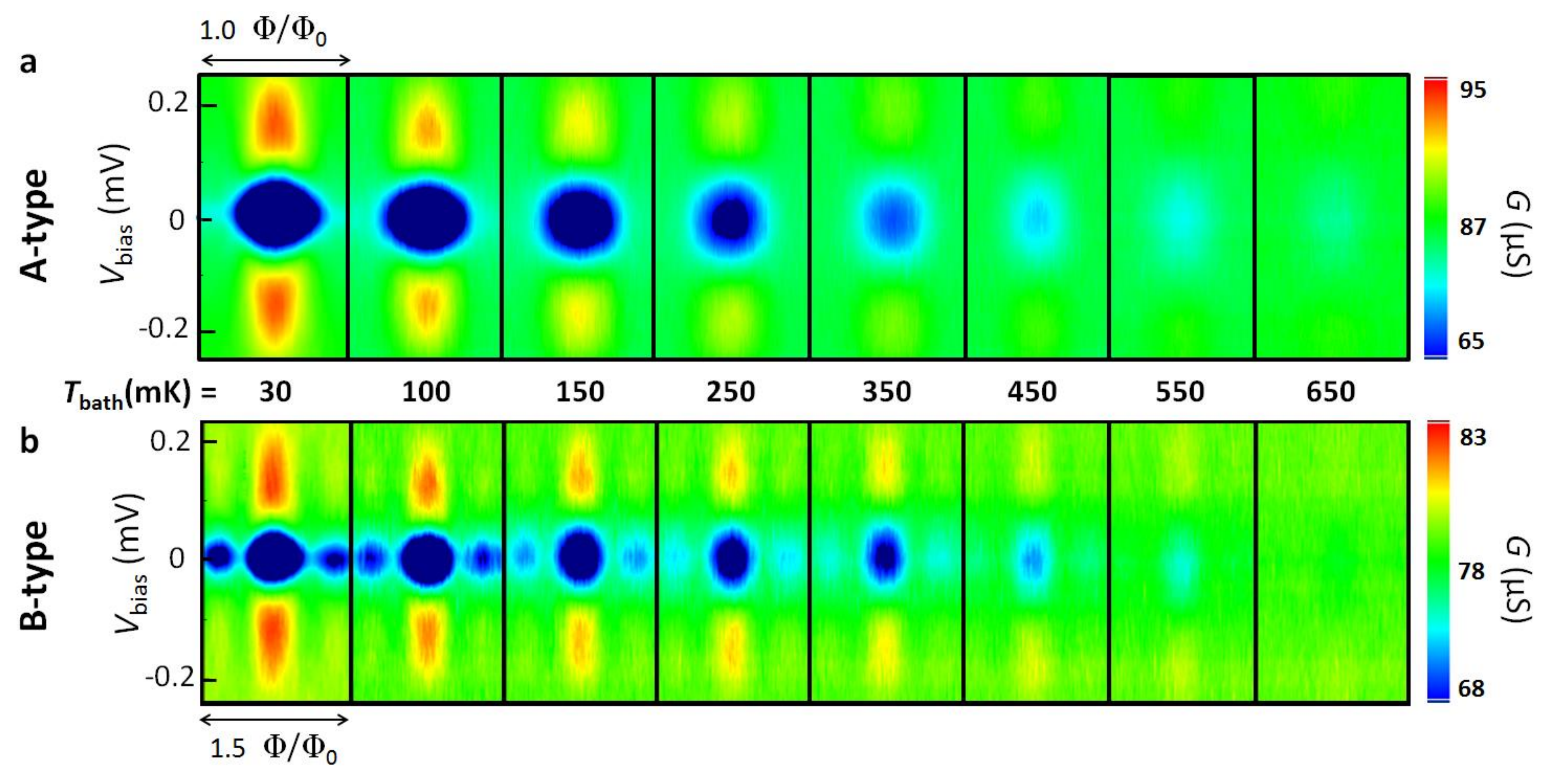}
  \caption{
  \label{Tdep}
  \textbf{Temperature evolution of the topological transitions.} 
  \textbf{a},\textbf{b}~Color plot of the tunneling conductance versus bias voltage and magnetic flux \textit{G}(\textit{V}$_{\text{bias}}$,$\Phi$) measured at several bath temperatures $T_{\text{bath}}$ for the A-type (\textbf{a}) and the B-type (\textbf{b}) interferometer. The double-minigap feature of B-type interferometer is clearly observable up to $\sim$300~mK whereas the magnetic flux-dependent behaviour of the conductance persists up to $\sim$500~mK in both type of $\omega$-SQUIPTs.}
\end{figure*}
Here we report the realization of the first phase-tunable three-terminal JJ interferometer based on proximity effect, in the following referred to as the  $\omega$-superconducting quantum interference proximity transistor ($\omega$-SQUIPT) due to its characteristic shape (see Fig.~\ref{sketch}). 
As shown below, this hybrid multi-terminal geometry allows to access a unique class of topologies defined in the (two-dimensional) domain of the superconducting phases thereby realizing a  \emph{Josephson topological material}.

The $\omega$-SQUIPT is fabricated by electron-beam lithography, three-angle shadow-mask evaporation of metals, and in situ oxidation (see Methods for further details). 
It essentially consists of three different parts, as shown in Fig.~\ref{sketch}a. 
Aluminum (Al) is used to form the superconducting double-ring structure consisting of three leads converging into a nanosized T-shaped copper (Cu) weak-link. 
A N electrode made of Al$_{0.98}$Mn$_{0.02}$ is tunnel coupled to the center of the T-shaped region, and allows to probe the weak-link DoS  through tunneling conductance measurements.
Specifically, at finite temperature the tunneling conductance ($G$) of the interferometer, measured according to the scheme presented in Fig.~\ref{sketch}b, is given by 
\begin{equation}
\label{eqG}
G(\textit{\text{V}},\Phi_L,\Phi_R)=\frac{1}{eR_{T}}\;\,\int_{-\infty}^{\infty}\;\, dE\;\, \mathcal{\bar N}_{N}(E,\Phi_L,\Phi_R)
\dfrac{\partial\mathcal{F}_0(E-e\textit{\text{V}})}{\partial V},\nonumber
 \end{equation}
where $\mathcal{\bar N}_{N}(E,\Phi_L,\Phi_R)$ is the DoS in the T-shaped region averaged over the probing junction area, $V$ is the voltage drop across the tunnel junction,  $\Phi_{L}$ and $\Phi_{R}$ are the magnetic fluxes piercing the left and right loop, respectively, and $\mathcal{F}_0$ is the Fermi-Dirac energy distribution function. 
Moreover, $R_T$ is the normal-state resistance of the junction and $e$ denotes the electron charge.

Two different families of $\omega$-SQUIPTs were fabricated and characterized: the A-type interferometer (shown in figure~\ref{sketch}a), and the B-type one which is less symmetric with respect to the length of the arms forming the T-shaped region, and to the resistances of the three Cu/Al interfaces.
These two families of interferometers allow to access different topologies of the electron states in the weak-link by tuning the external magnetic flux, as shown in Fig~\ref{Topology}.
In analogy with conventional topological insulators we assign to the gapped (insulating) states of the weak-link a topological index defined by a couple of discrete numbers ($n_L,n_R$) obtained via the line integral of the superconducting-phase  gradient ($\vec{\nabla} \varphi({\bf r})$) over the left and right S loops of the interferometer (see Fig~\ref{Topology}a).
This index, similarly to the Chern number, identifies \emph{univocally}  different topologies of the gapped states. 
These are represented by the blue regions in the plots of the DoS at the Fermi energy calculated in Fig.~\ref{Topology}b,c as a function of the magnetic fluxes $\Phi_L$ and $\Phi_R$ for 
the A-type and B-type $\omega$-SQUIPT, respectively.
In full analogy with the physics of topological insulators, the interface between two regions characterized by different topologies is a conductive state represented by the green (gapless) area appearing in the plots of the DoS.
The main difference between the diagrams of the two families lies in the shape of the gapped blue zones, which is characterized by a pronounced skewing in  B-type interferometer (Fig.~\ref{Topology}c). 
This allows the crossing between the four different topologies (00,10,01,11) and the dotted line defined by the relation $\Phi_L = \Phi_R$, the latter representing the phase space accessible in our experiment.
Therefore, in A-type interferometer a single topological transition between 00 and  11 gapped states is expected to occur within a flux quantum $\Phi_{0}$ ($\Phi_{0}\simeq 2\times 10^{-15}$ Wb). By contrast, in B-type $\omega$-SQUIPTs all the four different topologies can be explored within the same magnetic-flux interval.

The peculiar \emph{topological protection} among different statescan be understood with the aid of the vectorial representation of the electron states in the weak-link~\cite{Nazarov_Circuit_1994} displayed in Fig~\ref{Topology}d,e.
In this picture the electron states are mapped through the position of an unitary vector in a hemisphere (see Fig.~\ref{Topology}f), with a \emph{longitudinal} angle ($\varphi$) representing the superconducting phase, and a \emph{latitudinal} angle ($\theta$) defining the degree of "superconductivity" of the weak-link [see Eq. (\ref{eq:norm}) and Supplementary Information].
In particular, the North pole ($\theta=0$, identified by a cross on top of each sphere) corresponds to a N-like behaviour (gapless state) whereas the equator line ($\theta=\pi /2$) represents an ideal superconductor (gapped state) with phase $\varphi$.
%
The four distinct topologies shown in Fig.~\ref{Topology}c,d corresponding to $(n_L\,n_R) = (00),(10),(01),(11)$ are then \emph{protected} since cannot be \emph{continuously} transformed into  another, without \emph{pulling} them out of the equator. 
In other words, each topological transition requires therefore the state vector (i.e., the electronic configuration of the T-shaped weak-link) to deviate from the equator resulting into a gapless state.
Figures~\ref{Topology}d,e sketch the evolution of these transitions between different topologies along the $\Phi_L = \Phi_R$ flux path (see Supplementary Information for further details).

These two distinct kinds of topological transitions are clearly observed in our interferometers.
Figure~\ref{TheEffect}a,b (left panels) shows the characteristic behaviour of the tunneling conductance ($G$) of the A-type $\omega$-SQUIPT measured at 30~mK versus bias voltage ($V_\text{bias}$) and magnetic flux ($\Phi=\Phi_L=\Phi_R$).
Experimental data are directly compared to the conductance spectra calculated from Eq.~(\ref{eqG}) for the T-shaped diffusive Cu weak-link (see Methods for details).
The comparison demonstrates the nearly optimal agreement between theory and experiment. 
Notably, two different regimes are clearly visible in the $G(V_\text{bias},\Phi)$ spectrum (see Fig. \ref{TheEffect}b): 
one occurring for  $|\Phi/\Phi_0|\lesssim1/4$ (with $\Phi_{0}$ periodicity) characterized by the presence of  a finite minigap, and another for $1/4 \lesssim\Phi/\Phi_0\lesssim3/4$ where the minigap is almost absent. 
The remaining small conductance dip, still visible around zero bias for $\Phi=0.5\Phi_0$ (see Fig. \ref{TheEffect}a), stems from the finite size of the probing junction which provides an averaged DoS of the weak-link over the tunneling area.
While the gapped regions provide a physics apparently similar to that observed in two-terminal JJs~\cite{Giazotto_Superconducting_2010,dambrosio_Normal_2015},
the existence of a gapless regime furnishes the experimental proof of a phase-induced topological transition from the 00 to the 11 topological state occurring in the three-terminal weak-link~\cite{Padurariu_Closing_2015,van_Heck_single_2014}. 
Between these two states the energy spectrum of Andreev bound states which determine the weak-link conductance crosses the (zero-energy) Fermi level within a continuos band of magnetic-flux values \cite{Padurariu_Closing_2015}. As a consequence of this energy crossing, a normal metal-like conductivity at zero-bias is recovered.
 
To further validate the above given theoretical interpretation we extend the analysis of the three-terminal interferometer to the more asymmetric B-type configuration.
A small change in the symmetry of the JJs forming the structure indeed drastically affects and modifies the characteristic behaviour of the $\omega$-SQUIPT (see Fig.~\ref{TheEffect}c,d), and a small gapped region appears also for $1/4 \lesssim\Phi/\Phi_0\lesssim3/4$.
As displayed in the phase diagram of Fig.~\ref{Topology}c, we ascribe the existence of this second insulating region to the presence of two additional topologies (10,01) which become accessible when the two lateral  S leads are more asymmetrically coupled to the T-shaped weak-link.
This complementary proof confirms the good agreement between the experiment and the topological picture given for the three-terminal JJ.

The robustness against bath temperature ($T_{\text{bath}}$) of the observed topological transitions for both types of interferometers is quantified in Fig.~\ref{Tdep}.
In particular, it shows the temperature evolution of the tunneling conductance spectra $G(V_\text{bias},\Phi)$ for the A- (pannel a) and B-type (pannel b) $\omega$-SQUIPTs. 
As expected, the impact of Fermi distribution at finite temperature manifests itself via a broadening of $G$, and through a gradual fade-out of its main features by increasing $T_{\text{bath}}$.
In particular, the topological gapped regions 10 and 01 are still clearly observable up to $\sim$~300~mK whereas the 00 and 11 topologies persist up to $\sim$~500~mK.
Access to higher temperatures is expected to occur by reducing the weak-link size \cite{Giazotto_Hybrid_2011} as well as by increasing the energy gap of the S loops~\cite{Ronzani_Micro-superconducting_2013}.

In summary, we have realized the first double-loop superconducting quantum interference proximity transistor with three superconducting terminals. 
This structure promotes an additional phase control as compared to the more conventional two-terminal geometry, and allows an exotic phase-engineering of the weak-link topology which manifests itself in the peculiar behaviour of the interferometer conductance. 
The gapped states induced in the weak-link realize a (two-dimensional) \emph{Josephson} topological material univocally classified by a topological index tunable  via an  external magnetic field.
Phase-tuning of the $\omega$-SQUIPTs allowed the access to a $\Phi_{0}$-periodic topological transition showing up
in a continuos magnetic-flux interval $1/4\lesssim \Phi/\Phi_{0}\lesssim 3/4$. 
Yet, increasing the asymmetry of the JJs forming the interferometer leads to the appearance of a second gapped state which reveals the presence of two additional topological phases induced in the weak-link, in full quantitative agreement with our topological model.
Being the first experimental demonstration of ultimate theories on multi-terminal Josephson junctions, this class of interferometers paves the avenue for future coherent nanoscale devices where quantum technology and fundamental research merge to be one \cite{padurariu_Spin_2012,Mourik_signatures_2012}. 
In this perspective, the $\omega$-SQUIPT could be easily combined with hybrid nanocircuits based on, e.g., semiconducting nanowires~\cite{Giazotto_Josephson_2011}, low dimensional systems~\cite{Deon_Proximity_2011}, graphene~\cite{Heersche_Bipolar_2007} or superconductors to enhance its functionalities.
Yet, supplying the interferometer with independent on-chip coils, which provide separate control over the two magnetic fluxes, would assure to master independently phase-biasing in left and right loop.
Finally, the addition of more superconducting terminals will increase the phase dimensionality of the weak-link electron states enabling further artificial topologies~\cite{Riwar_Multi-terminal_2015}.

\section*{Methods}
\textbf{Fabrication details and experimental set-up.} The $\omega$-SQUIPTs were fabricated with electron-beam lithography and three-angle shadow-mask evaporation of metals onto an oxidized Si wafer through a bilayer resist mask. The evaporations and oxidation were made in an ultra-high vacuum electron-beam evaporator, which allowed us to deposit first 15 nm of Al$_{0.98}$Mn$_{0.02}$ at an angle of 40$^{\circ}$ to form the N probe. 
Then the sample was exposed to 40~mTorr of O$_{2}$ for 5 minutes to realize the insulating layer of Al$_{0.98}$Mn$_{0.02}$Ox forming the tunnel barrier. Afterwards, the sample was tilted at an angle of 20$^{\circ}$, and 25~nm of Cu were deposited to realize the T-shaped N wire. 
Finally, 150~nm of Al were evaporated at an angle of $0^{\circ}$ to implement the double S ring with three terminals S$_{\text{L}}$, S$_{\text{C}}$, and S$_{\text{R}}$ (see Fig.\ref{sketch}b). 
To achieve full phase polarization the $\omega$-SQUIPT requires a superconducting double-loop with a cross section much larger than the T-shaped  N wire in order to reach the condition $\mathcal{L}^{R}\ll$ $\mathcal{L}^{WL}$, where $\mathcal{L}^{R(WL)}$ denotes the kinetic inductance of the ring (weak-link) \cite{le_sueur_phase_2008,Ronzani_Highly_2014,dambrosio_Normal_2015}.
The magneto-electric characterization of the interferometers was performed in a filtered He$^{3}$-He$^{4}$ dilution refrigerator at different temperatures ranging from 30~mK to 1.2~K.
Voltage and current were measured with standard room-temperature preamplifiers. 
A lock-in amplifier was used to reduce noise in the tunneling conductance measurements.
Five $\omega$-SQUIPTs have been fully characterized, two of which were of A-type and three of B-type.
\newline

\textbf{Theoretical model.} To model the $\omega$-SQUIPT we consider the 3-terminal T-shaped Josephson weak-link sketeched in Fig.\ref{sketch}b. 
It consists of three diffusive quasi-one dimensional N arms with lengths $L_{i}$ ($i=\text{L,C,R})$, each of them connected to a superconducting electrode $S_{i}$ with phase $\varphi_{i}$. In order to determine the density of states in the N arms we introduce the isotropic quasiclassical retarded Green functions $\hat{g}{}_{i}$ which are $2\times2$ matrices in the Nambu space. These functions satisfy in each of the N arms the Usadel equation~\citep{Usadel_Generalized_1970}
\begin{equation}
\partial_{x}\left(\hat{g}_{i}\partial_{x}\hat{g}_{i}\right)+\frac{i(E+i\delta)}{E_{_{i}}}\left[\hat{\tau}_{3},\hat{g}_{i}\right]=0\,,\label{eq:Usadel}
\end{equation}
where $x$ is the dimensionless coordinate normalized in each arm  by the respective length $L_{i}$, and $E_{i}=D/L_{i}^{2}$ is the Thouless energy of the $i$-th arm, $\hat{\tau}_3$ is the third Pauli matrix in the Nambu space, and $\delta$ takes into account the inelastic scattering in the N region. Equation (\ref{eq:Usadel}) is complemented by the normalization condition:
\begin{equation}
\hat{g}_{i}^{2}=1.\label{eq:norm}
\end{equation}
This condition allows to write the matrix Green  function in terms of an unit vector $\vec g$ such that  $\hat g=\vec g.\vec \tau$. Here $\vec \tau=(\hat \tau_1,\hat \tau_2,\hat \tau_3)$ is the vector of Pauli matrices in Nambu space, and $\vec g=(\cos\varphi \sin \theta, \sin \varphi \sin \theta, \cos \theta)$ (see the scheme shown in Fig. \ref{Topology}f where the angles $\theta$ and $\varphi$ are defined, and Supplementary Information for further details).

To describe each S/N interfaces we use the Nazarov's boundary condition \cite{Nazarov_Circuit_1994}
\begin{equation}
r_i\hat{g}_{i}\partial_{x}\hat{g}_{i}=\frac{2\left[\hat{g}_{i},\hat{G}_{i}\right]}{4+\tau\left(\left\{ \hat{g}_{i,}\hat{G}_{i}\right\} -2\right)}\;,\label{eq:bc0}
\end{equation}
where $\hat{G_{i}}=E/\sqrt{(E+i\delta)^{2}-|\Delta_{i}|^{2}}\hat{\tau_{3}}+i\hat \tau_{1}\hat \Delta_{i}/\sqrt{(E+i\delta)^{2}-|\Delta_{i}|^{2}}$ is the BCS Green's function of the $S_{i}$ electrode, $\hat\Delta_{i}=\Delta e^{i\varphi_{i}\hat \tau_3}$ is the superconducting order parameter, $r_i=G_{N_{i}}/G_{B_i}$ is the ratio between the conductance of each arm ($G_{N_{i}}$) and the barrier conductance, $G_{B_i}=G_{0}n_i\tau$, where $G_{0}$ is the quantum of conductance, and $n_i$ the number of each SN interface conducting channels that we assume to have the same transmissivity $\tau$. 
Furthermore, we set $\varphi_C=0$ and neglect the inductance of the superconducting loops so that we can define the phase differences $\varphi_L=2\pi\Phi_L/\Phi_0$ and $\varphi_R=-2\pi\Phi_R/\Phi_0$, where $\Phi_{L(R)}$ is the total magnetic flux through the left (right) loop area, $\Phi_0=h/2e$ is the flux quantum, and $e$ is the electron charge.
At the crossing point between the arms ($x=0$) we impose the continuity of the Green's functions and the conservation of the matrix current which translate into following conditions:
\begin{eqnarray}
\sum_{i=1}^{3}G_{N_i}\left.g_{i}\partial_{x}g_{i}\right|_{x=0} & = & 0\label{eq:bc1}\\
g_{1}(0)= & g_{2}(0)= & g_{3}(0).\label{eq:bc2}
\end{eqnarray}
By solving numerically the non-linear boundary problem defined by Eqs.~[(\ref{eq:Usadel})-(\ref{eq:bc2})], and from the solution $g(x)$ at the crossing point one determines the density of states $\mathcal{N}_N(x,E,\Phi_L,\Phi_R)=\frac{1}{2}\text{Tr}\left[\hat{\tau_{3}}\text{Re}[\hat{g}(x)]\right]$.
The tunneling conductance measured by the normal metal probe can be obtained by deriving with respect to the voltage $V$ the expression for the tunneling current 
\begin{eqnarray}
I(V,\Phi_L,\Phi_R)=\frac{1}{eR_T}\int^{\infty}_{-\infty} dE\mathcal{\bar N}_N(E,\Phi_L,\Phi_R)\nonumber\\
\times\left[\mathcal{F}_0(E-eV)-\mathcal{F}_0(E)\right],\nonumber
\end{eqnarray}
where $\mathcal{\bar N}_N(E,\Phi_L,\Phi_R)$ is the density  of states averaged over the probing junction area, $\mathcal{F}_0(E)$is the Fermi-Dirac distribution function, and $R_T$ the junction normal-state resistance.
For the calculations of the A-type $\omega$-SQUIPT we used the following parameters: $E_L=0.59\Delta_0$, $E_C=\Delta_0$, $E_R=0.89\Delta_0$ and $\Delta_0=190\,\mu$eV as the zero-temperature superconducting energy gap, $\tau=1$,$\delta=0.12\Delta_0$, $r_L=2.5,r_C=1.3,r_R=0.1$,$G_{N_R}/G_{N_C} = 0.625, G_{N_L}/G_{N_C} = 0.66$. For the B-type we set: $E_L=0.44\Delta_0$, $E_C=0.92\Delta_0$, $E_R=0.76\Delta_0$, $\tau=1$,$\delta=0.13\Delta_0$, $r_L=3,r_C=4.9,r_R=0.2$, $G_R/G_C = 0.575, G_L/G_C = 0.68$.  

\section*{Acknowledgements}
The European Research Council under the European Union's Seventh Framework Program (FP7/2007-2013)/ERC Grant agreement No. 615187-COMANCHE and MIUR-FIRB2013 -- Project Coca (Grant No.~RBFR1379UX) are acknowledged for partial financial support.
The work of E.S. is funded by the Marie Curie Individual Fellowship MSCA-IFEF-ST No. 660532-SuperMag. 
The work of F.S.B was partially supported by the Spanish Ministerio de Economia y Competitividad under Project 
No. FIS2014-55987-P. 
\section*{Author contributions}
S.D. fabricated the samples. 
E.S. and F.V. performed the measurements.
F.V. analyzed the data, and carried out the simulations.
F.S.B developed the numerical code to calculate the conductance spectra.
Y.V.N. developed the theory of the Josephson topological states.
F.G. conceived the experiment. 
All authors discussed the results and their implications equally at all stages, and all the authors wrote the manuscript.

\section*{Competing financial interests}
The authors declare no competing financial interests.


\newpage~
\newpage~
\section*{Supplementary Information}
\subsection*{Topological considerations for semiclassical proximized nanostructures}
In this section we define topological indices for superconducting states in nanostructures with a typical size largely exceeding  the electron wavelength, which is certainly the case of the present interferometers. In this case, the superconducting state in each point ${\bf r}$ of the normal part of the structure is characterized by a traceless $2 \times 2$ matrix Green function that generally depends on energy, $\hat{g}(\epsilon,{\bf r})$ and satisfies the condition~\cite{Nazarov_Quantum_2009} $\hat{g}^2=\hat{1}$ [{\it cf.} Eq. (\ref{eq:norm})]. At $\epsilon=0$, $\hat{g}$ can be mapped on a unit vector at a hemisphere~\cite{Nazarov_Circuit_1994}, $\hat{g} \to \vec{g} \to  (\cos\varphi \sin \theta, \sin \varphi \sin \theta, \cos \theta)$, $\varphi,\theta$ being the longitude and latitude, respectively, $-\pi \le \varphi \le \pi$, $0\le \theta \le \pi/2$ (see Fig. \ref{Topology}f). The states at the equator, $\theta=\pi/2$, are  superconducting and gapped, like those in the superconducting leads where $\varphi$ equals to the  phase of the superconducting pairing potential $\Delta({\bf r})$.  The deviation of the vector  from the equator $\theta({\bf r}) \ne \pi/2$ gives rise to finite density of states at zero energy  ($\epsilon=0$). This density is given by  $\nu({\bf r}) = \nu_0 \cos\theta$,  where $\nu_0$ is  the density of states in the normal metal that corresponds to the North pole of the hemisphere ($\theta=0$). 
\begin{figure}[b!]
\centerline{\includegraphics[width=0.8\columnwidth]{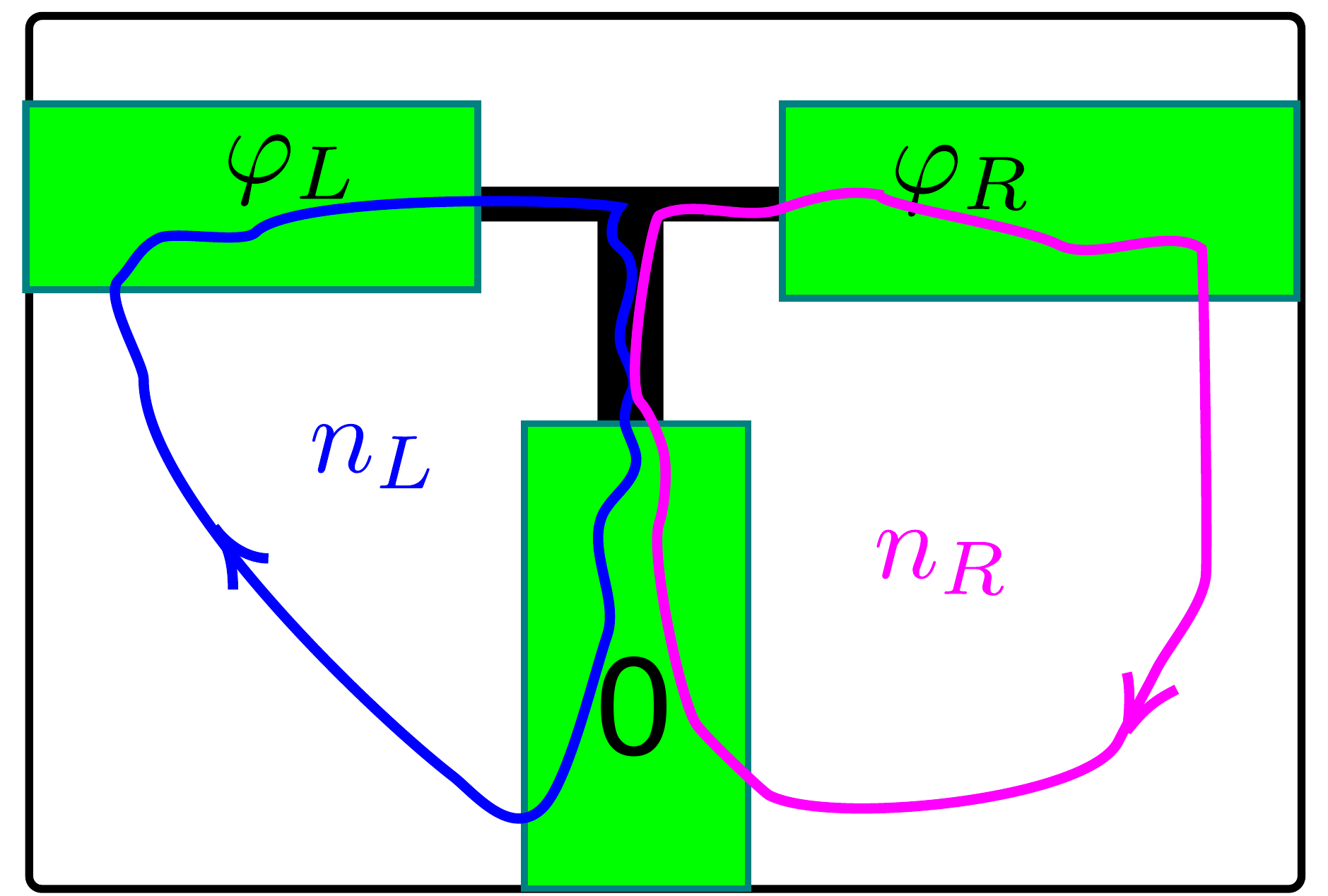}}
\caption{
\label{contours}
Two contours defining the topological numbers $n_{L,R}$.}
\end{figure}
The values of $\vec{g}$ at zero energy are determined by boundary conditions set on the terminals of the nanostructure. If all terminals are superconducting and are thus situated at the equator, they pull $\vec{g}$ to the equator too, while no agent favors the deviations towards the North pole. It is therefore natural to have the proximity gap overall the structure, with no apparent reason for a gapless state. We argue that the gapless state  in a proximity nanostructure and observed in our experiment arises from topological reasons. As we discuss below, one may ascribe topological indexes to the gapped states. The phases differing in the indexes occupy different regions of the parameter space. Owing to continuity, these regions must be separated by the regions of the gapless phase.

One can define the topological numbers for gapped states with well-defined $\varphi({\bf r})$ in a way that goes back to the discovery of the flux quantization~\cite{Deaver_Experimental_1961,Doll_Experimental_1961} by integrating the gradient of $\varphi({\bf r})$ over a closed contour, 
\begin{equation}
n = \frac{1}{2\pi}\oint d{\bf r} \vec{\nabla} \varphi({\bf r})\; .
\label{int_n}
\end{equation}
The number $n$ is an  integer by construction. The definition can be expanded if there are small discontinuities in $\varphi({\bf r})$, for instance, at  tunnel barriers in the nanostructure. In this case, one adds to the integral the value of the phase jump projected on $(-\pi,\pi)$ interval as it is done in the description of Josephson arrays \cite{Fazio_Quantum_2001,Nazarov_Quantum_2009}.  
Within the normal-metal part of the structure there are no phase singularities, and hence  if the contour is within the proximized structure the integral (\ref{int_n}) vanishes.  

For our device, we may define two topological numbers extending the contours over the left or right loops, as shown in Fig. \ref{contours}.
\begin{figure}[t!]
\centerline{\includegraphics[width=0.8\columnwidth]{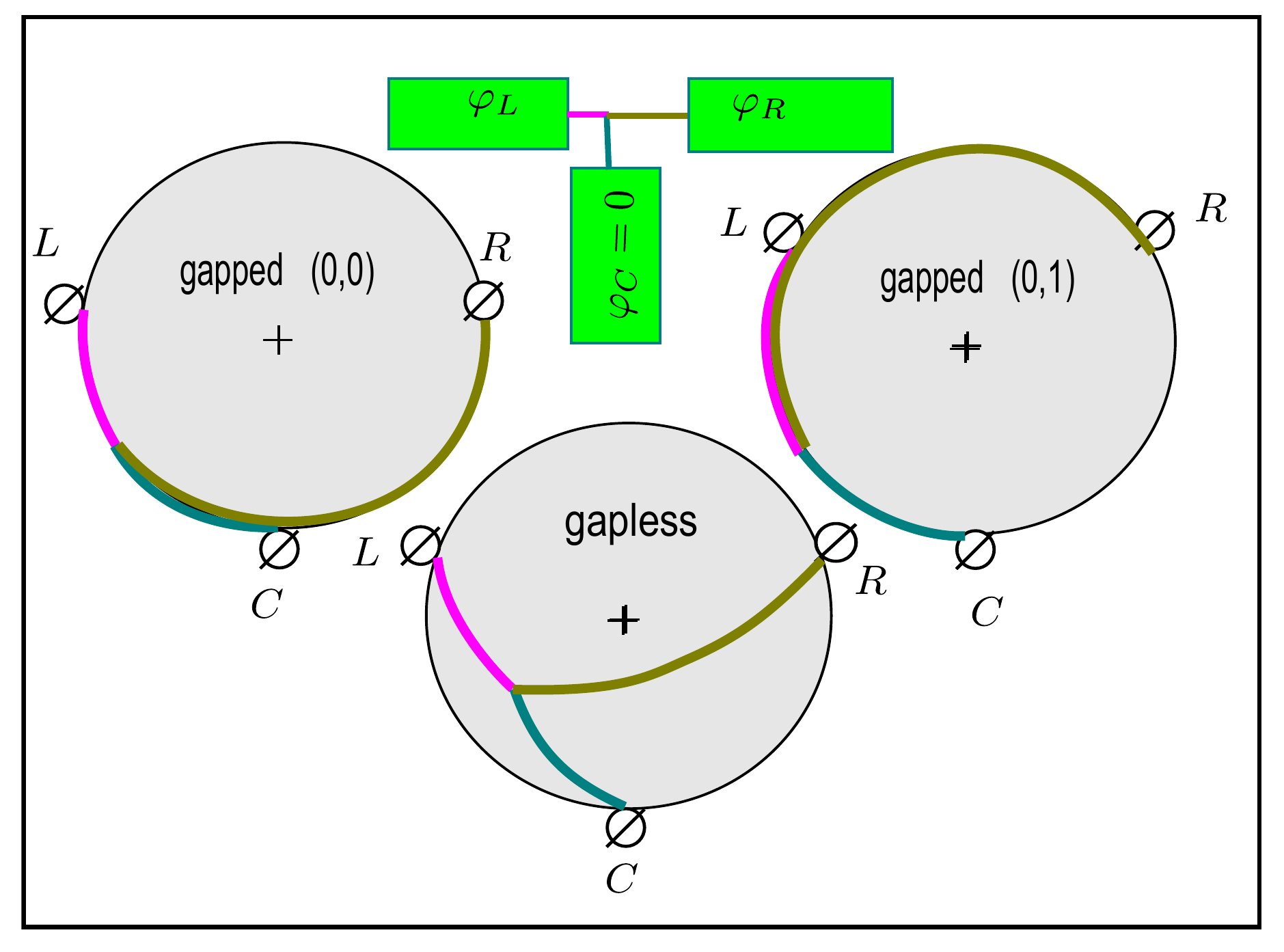}}
\caption{
\label{hemisphere}
Sketch of $\vec{g}({\bf r})$ dependence for three 1d wires given by distinct colours. Top views of the hemisphere for gapped states $(0,0)$ and $(0,1)$, and the  separating gapless state where $\vec{g}({\bf r})$ deviates from the equator. }
\end{figure}
There are two equivalent ways to incorporate the phase jumps $\varphi_L$,$-\varphi_R$ into these topological numbers. One can project these phase differences on $(-\pi,\pi)$ interval. In this case, the numbers are obviously periodic in fluxes $\Phi_{L,R}$ in both loops. There is, however, an inconvenience: a number jumps by $1$ any time the corresponding phase difference passes $\pi$, while this jump does not signify any change in the gapped superconducting state of the junction. It is therefore  more consistent to replace the phase differences with the actual values of the flux in the loop,   
\begin{equation}
\varphi_L, - \varphi_R \to 2\pi\frac{\Phi_{L}}{\Phi_0},2\pi\frac{\Phi_{R}}{\Phi_0} \; .
\end{equation}
In this way, the topological numbers do not jump within a domain of a given gapped state. 
However, they are now obviously not periodic in flux. We stick to the second definition.

\begin{figure*}[t!]
\includegraphics[width=\textwidth]{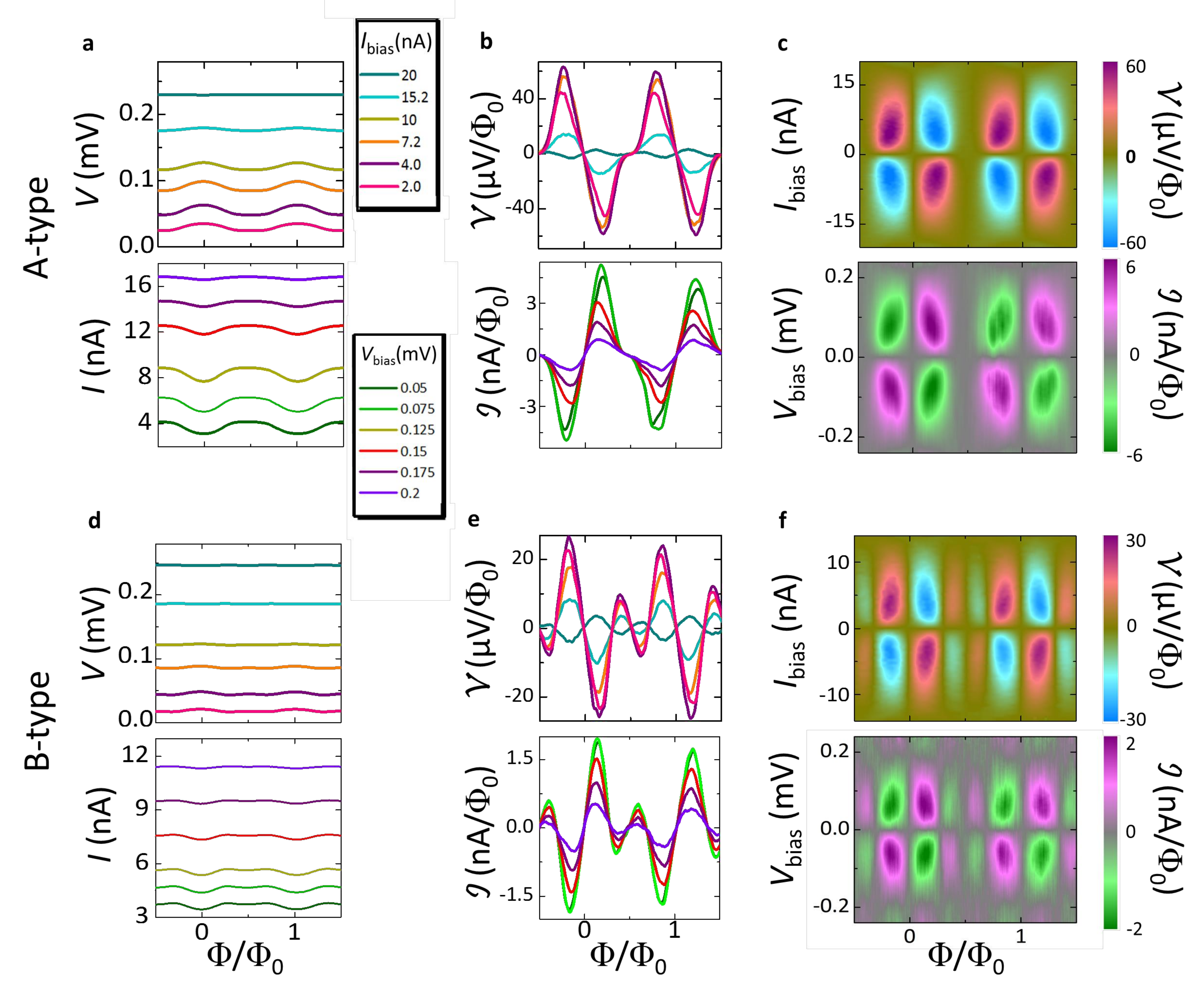}
  \caption{
  \label{Performance}
  \textbf{Magnetic-flux sensitivity of the A- and B-type $\omega$-SQUIPTs.}
  \textbf{a},\textbf{d},~Flux-modulated voltage drop measured at different current bias \textit{V}(\textit{I}$_{\text{bias}}$,$\Phi$) (on the top) and flux-modulated current at different voltage bias \textit{I}(\textit{V}$_{\text{bias}}$,$\Phi$)(on the bottom) for the A-type and B-type $\omega$-SQUIPT, respectively. 
  \textbf{b}, \textbf{e},~Flux-to-voltage transfer function ($\mathcal{V}=\partial V/\partial \Phi$) (on the top) and flux-to-current transfer function ($\mathcal{I}=\partial I/\partial \Phi$) (on the bottom) obtained from measurements of panel \textbf{a},\textbf{d}. 
  \textbf{c}, \textbf{f},~Color plots showing the flux-to-voltage transfer function vs current  and flux, $\mathcal{V}$(\textit{I}$_{\text{bias}}$,$\Phi$) (on the top), and flux-to-current transfer function  vs voltage and flux, $\mathcal{I}$(\textit{V}$_{\text{bias}}$,$\Phi$) (on the bottom), for the A-type (panel \textbf{c}) and B-type (panel \textbf{f}) $\omega$-SQUIPT, respectively. All the measurements are taken at 30~mK.}
\end{figure*}
If we consider an  elementary cell in flux space we find 4 states of distinct topology with $(n_L,n_R) = (0,0),(1,0),(0,1),(1,1)$.  They are topological protected in the sense that lying at the equator these gapped configurations of $g({\bf r})$ cannot be continuously transformed to one another.  The actual transition requires  $\vec{g}({\bf r})$ to deviate from the equator and this results in a gapless state. 

Let us illustrate this with exemplary nanostructure made of three quasi-1dimensional wires connected to the corresponding terminals $L,C,R$ and joining in the same point (see Figure \ref{hemisphere}). The three wires are represented  by three distinct colors.

We plot the values of $\vec{g}({\bf r})$ across the wires on the top view of the hemisphere. The corresponding curves start in respective terminals and join together in some point. For gapped states, all curves lie at the equator. A qualitative difference between the $(0,0)$ and $(0,1)$ states  is the position of the olive curve: it is on different sides of the equator. The continuous transition between these two configurations require the olive curve to go up in latitude while  pulling the other curves. This leads to  a finite density of states in the weak-link.

\begin{figure}[t!]
\includegraphics[width=0.98\columnwidth]{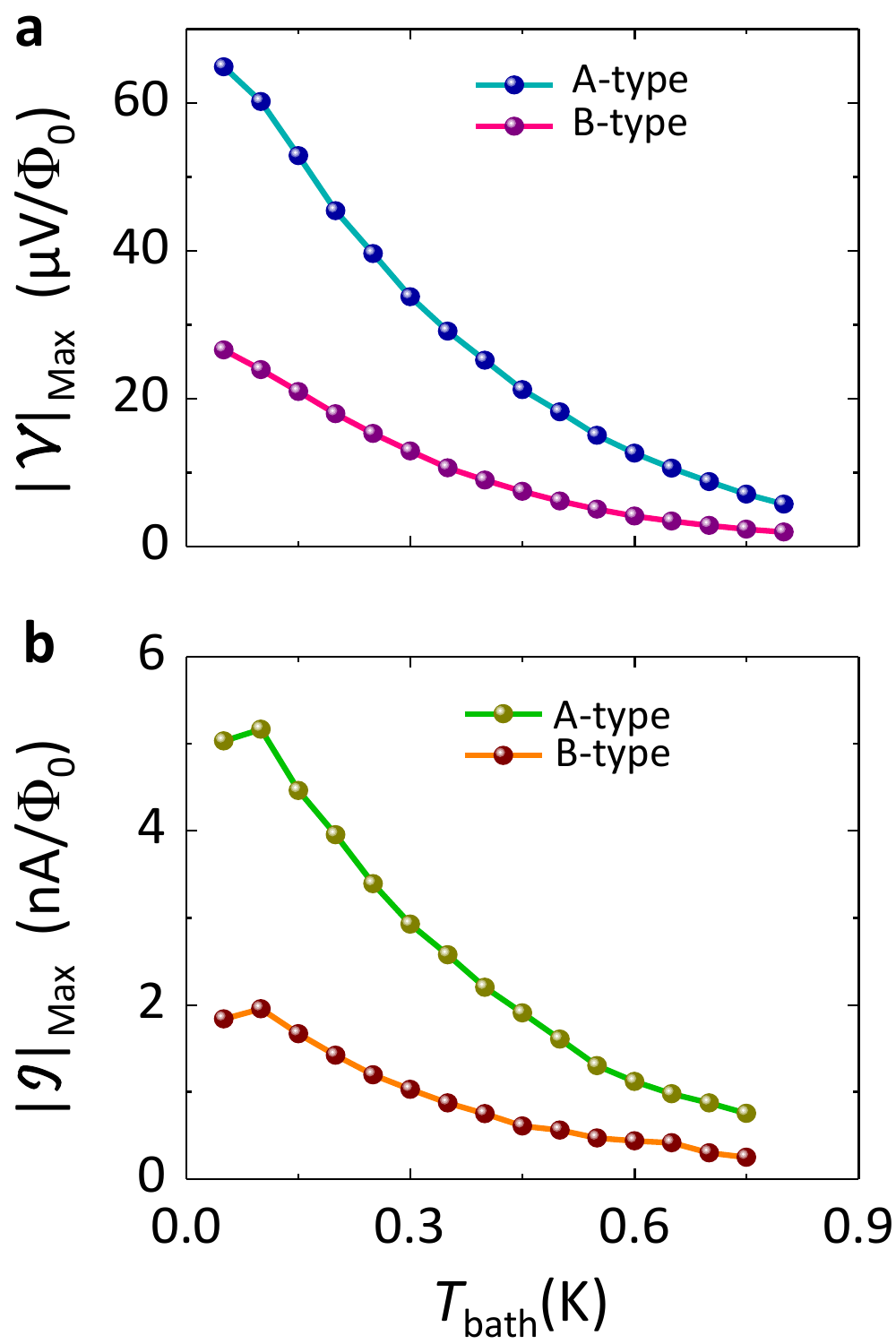}
  \caption{
  \label{PerformanceT}
  \textbf{Temperature behavior of the transfer functions of the A- and B-type $\omega$-SQUIPTs.}
 \textbf{a,b}~Temperature dependence of the maximum value of the flux-to-voltage (top panel) and flux-to-current (bottom panel) transfer function for the A- and B-type $\omega$-SQUIPT. The lines connecting the experimental data in both panels are guides to the eye.
}
\end{figure}

\begin{table}[th!]
\begin{center}
\begin{tabular}{l c c c c c}
\hline
\hline
      & $d$ & $R_{T}$ & $|\mathcal{I}|_{\text{Max}}$ & $|\mathcal{V}|_{\text{Max}}$ & Ref. no. \\
 Interferometer & (nm) & (k$\Omega$) & (nA/$\Phi_{0}$) & ($\mu$V/$\Phi_{0}$)& \\
 \hline
 $\omega$-SQUIPT & 450 & 6.5 & 5 & 60 & This paper\\

 N-SQUIPT & 160 & 33  & 12 & 450 & \cite{dambrosio_Normal_2015}\\

 S-SQUIPT & 1000 & 50 & - & 50 & \cite{Giazotto_Superconducting_2010}\\

 S-SQUIPT & 140 & 55 & 100 & 3000 & \cite{Ronzani_Highly_2014}\\
 \hline
 \hline
\end{tabular}
\end{center}
\caption{\label{tab1} Main parameters and performance of four different SQUIPT interferometers measured at $T_{\text{bath}}\sim 30$ mK. The symbol $d$ is used to identify the interelectrode distance between two superconducting leads, and $R_{T}$ is the normal-state resistance of the tunnel probe. $|\mathcal{I}|_{\text{Max}}$ and $|\mathcal{V}|_{\text{Max}}$ denote the maximum absolute value of the flux-to-current and
flux-to-voltage transfer functions, respectively. 
}
\end{table}

\subsection*{Performance as a magnetometer}
Besides the exotic topologies accessible with multi-terminal JJs it is also interesting to investigate the flux sensitivity of this novel double-loop geometry, and to compare it to that of existing proximity-based state-of-the-art superconducting interferometers.
A complete quantification of the flux sensitivity for the two $\omega$-SQUIPTs is reported Fig.~\ref{Performance} for both  current and voltage polarization.
Figure~\ref{Performance}a shows the current and voltage drop measured in the A-type interferometer at different bias from which the to flux-to-voltage ($\mathcal{V}$) and flux-to-current ($\mathcal{I}$) transfer functions are evaluated (see Fig.~\ref{Performance}b,c).
In particular, the modulations reach peak-to-peak amplitudes as large as $\delta$\textit{V}$\simeq$ 25 $\mu$V, and $\delta$\textit{I}$\simeq$ 2 nA.  
Moreover, the transfer functions obtain a maximum value up to $|\mathcal{V}|_{\text{Max}} \simeq$ 63 $\mu$V/$\Phi_{0}$, and  $|\mathcal{I}|_{\text{Max}}\simeq$ 5 nA/$\Phi_{0}$ at $\Phi\simeq\Phi_0/8$, and show the expected $\Phi_{0}$-periodicity. 
A similar characterization for the B-type $\omega$-SQUIPT is shown in Fig.~\ref{Performance}d,e,f. 
In general, it turns out that the performance of this interferometer are degraded as compared to the A-type. Indeed, the B-type $\omega$-SQUIPT reaches only modulations up to $\delta$\textit{V}$\simeq$ 5 $\mu$V, and $\delta$\textit{I}$\simeq$ 0.5 nA, then $|\mathcal{V}|_{\text{Max}} \simeq$ 28 $\mu$V/$\Phi_{0}$ and $|\mathcal{I}|_{\text{Max}}\simeq$ 2 nA/$\Phi_{0}$.

The best achieved performance of the $\omega$-SQUIPTs are compared in Table~\ref{tab1}  with those obtained in previous two-terminal SQUIPTs realized with a N~\cite{dambrosio_Normal_2015} or a superconducting probe~\cite{Giazotto_Superconducting_2010,Ronzani_Highly_2014}. 
This  comparison shows that the sensitivity of the three-terminal interferometer, although no optimal, is better than the sensitivity obtained in the first generation of two-terminal SQUIPTs~\cite{Giazotto_Superconducting_2010}, but it is somewhat lower than that of an optimized SQUIPT interferometer~\cite{Ronzani_Highly_2014} due to the smaller size of the weak-link, and thanks to the use of a S probe.

The impact of bath temperature ($T_{\text{bath}}$) on the transfer functions of both types of interferometers is displayed in Fig. \ref{PerformanceT}. In particular, both the flux-to-voltage (panel a) and the flux-to-current (panel b) transfer functions monotonically decay by increasing temperature, vanishing around $\sim 800$mK. 

Finally, it is worthwhile to mention that the double loop geometry of the $\omega$-SQUIPT may offer interesting perspectives if used as a flux-gradiometer at the nanoscale, owing to the strong dependence of the DoS in the weak-link on the difference between the two magnetic fluxes, ($\Phi_L-\Phi_R$)~\cite{Padurariu_Closing_2015}.

\end{document}